\documentclass[11pt,journal,compsoc]{IEEEtran}
\usepackage{graphicx}
\usepackage{algorithmic}
\usepackage{algorithm}
\usepackage{multicol}
\usepackage{amsmath}
\usepackage{amssymb}

\begin{document}
%
\title{Analysis of Distributed Algorithms for Big-data}

\author{\IEEEauthorblockN{ \textsuperscript{1.} Rajendra Purohit, K R Chowdhary   \textsuperscript{2.} S D Purohit}\\
\IEEEauthorblockA{1. Dept. of Computer Science  and Engineering,
Jodhpur Institute of Engineeering and Technology, Jodhpur \\
2. Dept. of HEAS (Mathematics), Rajasthan Technical University, Kota\\
Email: 1. rajendra.purohit@jietjodhpur.ac.in, kr.chowdhary@acm.org, 2. sdpurohit@rtu.ac.in }}

\maketitle


\begin{abstract}
The parallel and distributed processing are becoming de facto industry standard, and a large part of the current research is targeted on how to make computing scalable and distributed, dynamically, without allocating the resources on permanent basis. The present article focuses on the study and performance of distributed and parallel algorithms their file systems, to achieve scalability at local level (OpenMP platform), and at global level where computing and file systems are distributed. Various applications, algorithms,file systems have been used to demonstrate the areas, and their performance studies have been presented. The systems and applications chosen here are of open-source nature, due to their wider applicability. 
\end{abstract}




\IEEEpeerreviewmaketitle

\section{Introduction}

\IEEEPARstart{D}{ue} to the fast growth of the Internet of (IoT), it is expected that by 2029, more than 20 billion items, from smartphones to wearable and simple monitors, will be connected to the Internet ~\cite{zhu2019applications}. Distributed Ledger Technologies (DLTs), which are new and have been used to solve the problem of communication in distributed systems, link these devices together. Till recent, the only practicable solution to problems involving significant amounts of data required the computations to be distributed  across many machines. In conventional parallel programming model (e.g., MPI), the developer is responsible for managing concurrency explicitly. The MapReduce solution is appealing because its functional abstraction helps in providing a paradigm that is easy to grasp, for designing of scalable and distributed algorithms~\cite{lin2009brute}. The MapReduce algorithm is based on the fact that most information processing tasks have the same fundamental structure (e.g., Web pages) to generate partial results which are later aggregated. The data units in MapReduce are stored in the local disks of computers of a cluster that uses a distributed file system. The Hadoop which is an open-source implementation of the MapReduce programming paradigm, and HDFS (Hadoop Distributed File System) -- an open-source distributed file system that provides the fundamental storage substrate, were used in the experiments conducted as part of this work.

\section{Distributed Algorithms}

From the beginning of distributed computing, the study of scheduling strategies for parallel processing has remained the most active areas of research. Utilizing scheduling heuristics which approximate an optimal schedule has remained a prevailing practice. However, it is impossible to compare the efficacy of vaious heuristics using analytical methods. One of options is to conduct back-to-back tests on actual platforms. This is possible on tightly coupled platforms, however, it is not possible on modern distributed platforms, e.g., those which use Grid based processing, due to the labor-intensive nature of the process and the inability to replicate experiments. Thus, the solution is to opt for the simulation approach, which not only allows us to replicate the experiments but also permits us to investigate a vast array of platforms and applications. The works presented in this article make use of the MapReduce and Hadoop distributed framework, which facilitates the simulation of distributed computing applications utilizing numerous algorithms. 

There are in fact many network simulators, like, NS, DaSSF, or OMNeT++, all of them concentrate on simulation of packets traveling in a network instead of on the network behavior as experienced by the application. In fact, none of existing simulation framework satisfy the requirements of distributed system other than MapReduce, which is open source, and allows for simulation of arbitrary performance fluctuations such as those common for real resources, due to the background load.

\section{MapReduce Architecture}

Two systems became common in response to the data explosion: 1) the Google file system (GFS), and 2) MapReduce. The former was a pragmatic solution to managing exabyte-scale data using commodity hardware, whereas the latter was an implementation of a time-tested design pattern applied to massively parallel processing on commodity machinery.

In the open source community, \textit{Hadoop} subsequently consisted of  Hadoop Distributed File System (HDFS) and the Hadoop MapReduce (HMR). However, it is a MapReduce system at its essence. The code is converted into \textit{map} and \textit{reduce} tasks, which are executed by Hadoop.

The MapReduce architecture is based on a loosely coupled design. The pre-processing engine of MapReduce is not dependent on its storage architecture, due to this the processing and storage layers can be scaled independent of each other. Its storage system typically consists of a Distributed File System (DFS), like, the Google File System (GFS)\cite{ghemawat2003google} and and that used in Hadoop Distributed File System (HDFS)~\cite{5496972}. The HDFS  is a Java implementation of Google File System. Based on the DFS partitioning strategy, the data units are divided into equal-sized chunks and distributed across a cluster of processors. Each data segment is input to a \textit{mapper}. Thus, if data-set is divided into $k$ pieces, the MapReduce will generate $k$ number of mappers to process the data.

\begin{figure}[!ht]
\centering
\includegraphics[scale=0.18]{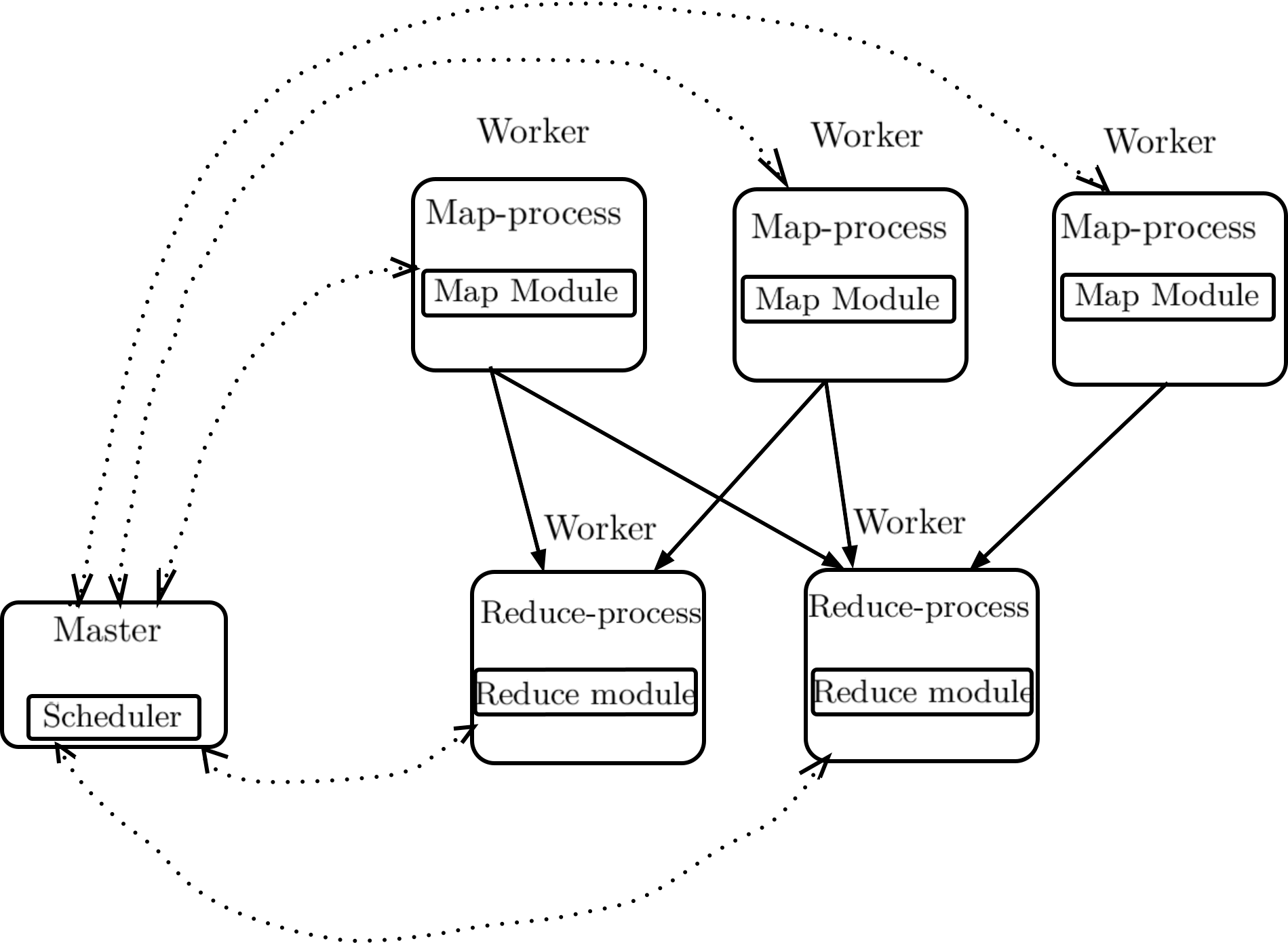}
\caption{MapReduce Architecture}
\label{fig:archmapred}
\end{figure}

The MapReduce processing engine comprises two types of nodes: 1. \textit{master nodes} and, 2. \textit{worker nodes} (Fig.~\ref{fig:archmapred}). The master node controls the execution flow of duties at the worker nodes through a scheduler module. Each worker node is in-charge of a map reduce operation. Each map worker node consists of a Input module, Map process,  Map module, combine module, and partition module, whereas each Reduce worker node consists of the Reduce process, Group process, Reduce module, and Output Module.

The scheduler distributes map and reduce jobs to worker nodes in a cluster, based on the criteria of locality of the data, current state of the network, and other variables. The Map module will read through a data chunk and then calls the user-defined map function to handle the data that was input. Once the intermediate results are generated, which are a set of key/value pairs, it arranges the results according to the partition keys, sorts the tuples within each partition, and notifies the master about positions of the results.

Once notified by the master, the Reduce module retrieves data from the mappers. After obtaining the intermediate results from the mappers, the reducer integrates the data by keys, and all the values with the same key are regrouped. Each key/value pair is then subjected to the user-defined function, and the results are sent to output to HDFS.

Given its goal of scalability across a large number of processing nodes, a MapReduce system must efficiently support fault tolerance. When any of the map or reduce task fails, a new task gets created on a different machine that will resume and rerun the unsuccessful task. Since the results are locally stored in the mapper, even a map task that has been accomplished, must be executed again in the event of a failed node. However, since the reducer retains the results in DFS, it is not necessary to re-execute a completed reduce task when a node fails.\medskip
  
\noindent  
{\bf MapReduce Applications}. One of the important application of MapReduce in the area of big-data is, indexing search-engine's index, to be later used in web search by Google and other search engines. The indexing system takes input, a large set of documents retrieved by the crawling system, and have been stored as GFS (Glarysoft Split File) files. The indexer takes input from these files and produces Google index in a distributed fashion. The salient features of this indexing are: simpler indexing code -- due to the fact that distribution and parallelization is hidden within the MapReduce library. Apart from this, scalability is assured as new machines can be added dynamically, as part of the indexing cluster.

\section{MapReduce Algorithm}

Hadoop is a free, community-developed software framework for analyzing and processing large data-sets using the MapReduce algorithm on a cluster of distributed computing nodes. By using the MapReduce programming model, we can distribute and process large data-sets in parallel. Before MapReduce, the data set had to be partitioned, with each section being assigned to a different processor and the final results being integrated.

\subsection{Traditional Method}

\begin{figure}[!ht]
\centering
\includegraphics[width=8cm, height=7.5cm]{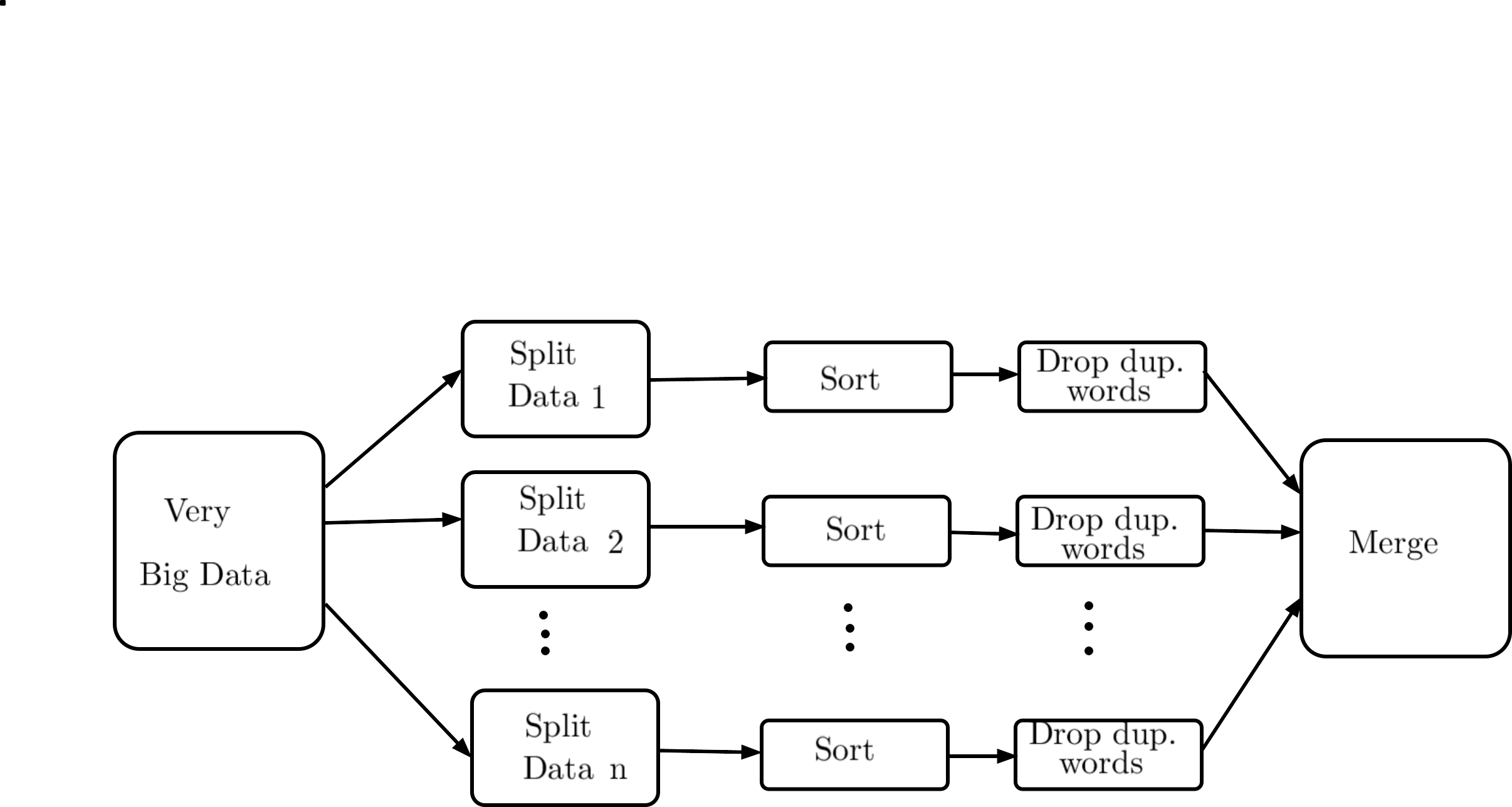}
\caption{Traditional processing}
\label{fig:alphsort}
\end{figure}

Utilizing the conventional approach, the algorithm that will be employed is as follows: The text should be divided into blocks that are about equivalent in size to the number of available computers or nodes. Subsequently, each individual element is concurrently arranged, eliminating any duplicate words within each partition, followed by the merging of the outcomes, which are subsequently transmitted to the output (Fig.~\ref{fig:alphsort}).

However, it has following drawbacks:

\begin{itemize}
\item if any of the machine delays the job, the whole whole is delayed;
\item if any of the machine fails, the whole work suffers (called reliability problem);
\item how to equally split the data?
\item fault tolerance needed (if any machine fails);
\item mechanism needed to produce output.
\end{itemize}

\subsection{MapReduce Algorithm}

\begin{figure}[!ht]
\centering
\includegraphics[width=8.5cm, height=7cm]{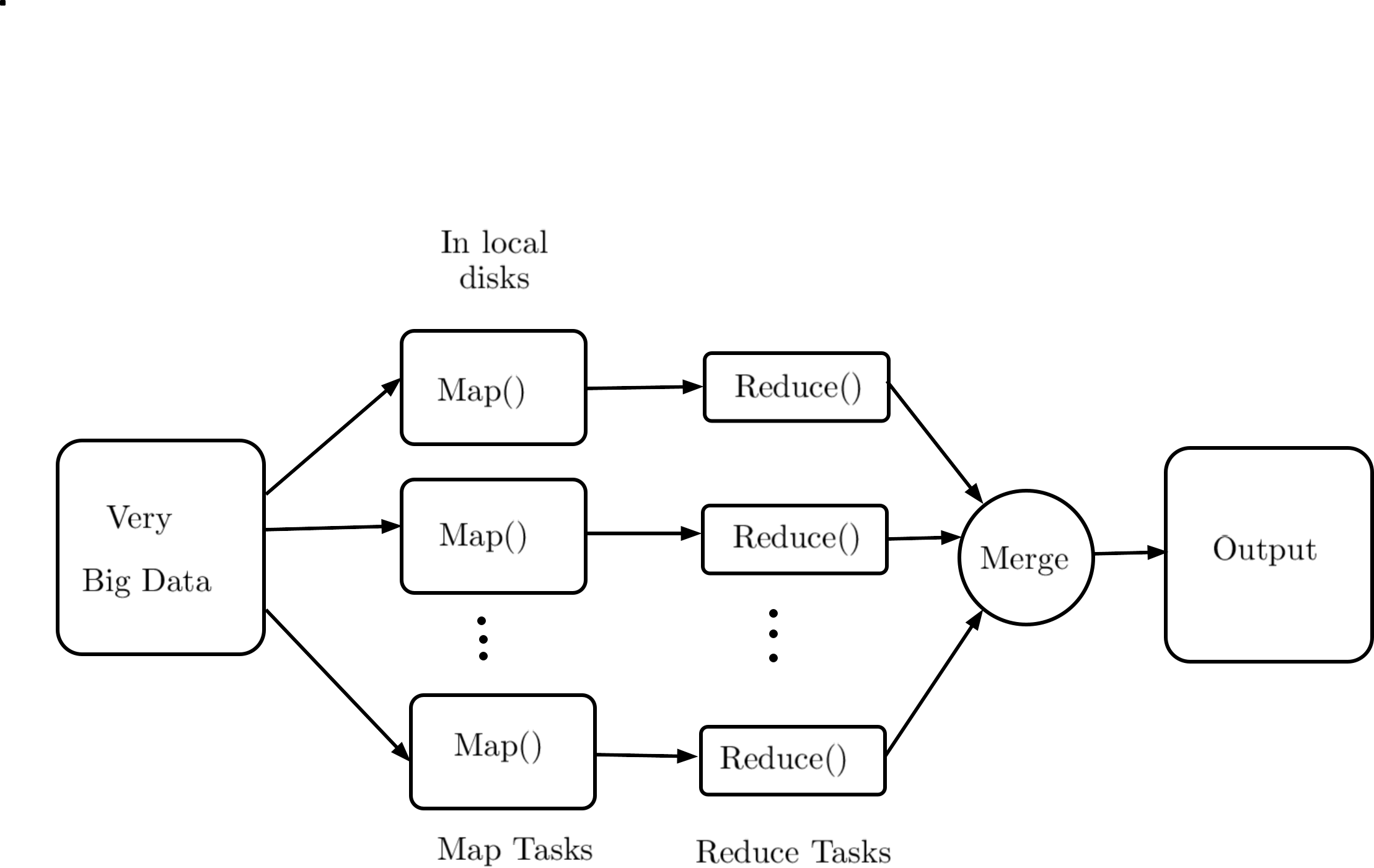}
\caption{MapReduce distributed algorithm}
\label{fig:mapredsol}
\end{figure}

A diagrammatic representation of the MapReduce algorithm shown in Fig.~\ref{fig:mapredsol} overcomes all these drawbacks.

The execution of MapReduce parallel-distributed processing is facilitated by three distinct classes \footnote{The term \textit{Class} originated from the Java programming language, where it is utilised for the implementation of MapReduce.}\medskip 

\noindent
{\bf Mapper Class}: The system is comprised of two separate categories of jobs, namely \textit{Map} and \textit{Reduce}. The reducer phase occurs subsequent to the completion of the map phase.  The \textit{key-value} pairs are produced as intermediate outcomes in the \textit{map()} operation by reading and processing blocks of data. The reducer blocks which are implemented using \textit{reduce()} function, receive key-value pairs from several \textit{map()} jobs. The reducer function condenses the intermediate data tuples, to be specific, the key-value pairs which are obtained through subsequent map() operations (Fig.~\ref{fig:aggrgt}) into a smaller collection of tuples or key-value pairs that becomes the final output.

The mapper class comprises: 1) input split, each of which produces a single block of data which is assigned as a single map task in the MapReduce program, and 2) a record-reader which interacts with the input split task,  and converts the obtained data in the form pf key-value pairs.\medskip

\noindent
{\bf Reducer Class}: Reducer classes are responsible for taking intermediate results from mapper classes and using them to create the final results, which are then written to the \textit{Hadoop Distributed File System} (HDFS).\medskip

\noindent 
{\bf Driver Class}: The driver class assumes the responsibility of configuring and initiating a MapReduce job to execute within the Hadoop framework.

The advantage of MapReduce \index{MapReduce} environment is that it can utilize commodity computer clusters for massively parallel processing of data. A MapReduce cluster may scale to thousands of nodes while remaining fault-tolerant. One of the primary benefits of this framework is in the utilization of a straightforward and robust programming paradigm. Additionally, it frees the application developer from all the intricate complexities of managing a distributed program, including problems with data distribution, scheduling, and fault tolerance ~\cite{sakr2013family}.

The distributed data processing system of MapReduce is based on the idea that parallel processing can be made easier by using a distributed computing platform. It has only two interfaces: \textit{map} and \textit{reduce}. Programmers make their own map and reduce functions, while the system is in charge of scheduling and coordinating the map and reduce tasks ~\cite{li2014distributed}.

\subsection{MapReduce Characteristics}

The MapReduce approach may be utilised to address "embarrassingly parallel" issues, which are those that take little or no effort to divide a work into a number of parallel but more manageable tasks. The MapReduce has applications in \textit{data mining}, \textit{data analytics}, and \textit{scientific computation}. The distinct qualities that it possesses are:\medskip

\noindent 
\textit{Flexible.} Programmers can write simple map and reduce functions to handle petabytes of data on thousands of machines without knowing how to run a MapReduce job in parallel.\medskip

\noindent
\textit{Scalable.}  One of the primary obstacles encountered in several contemporary applications is the ability to effectively handle and process the ever expanding quantities of data, which MapReduce can easily accommodate by allowing for data-parallel processing.\medskip  

\noindent
\textit{Efficient}. The MapReduce does not need loading of data into a database, hence saving the costs.\medskip

\noindent
\textit{Fault tolerance}. It refers to the ability of a system to continue functioning properly in the presence of faults or errors. In the MapReduce,  execution of each job involves the partitioning of tasks/jobs into numerous smaller units, which are subsequently distributed among multiple computers for processing. Failure of a task or machine is compensated by assigning the task to a machine capable of handling the burden. The job's input is stored within a distributed file system, which maintains many replicas to provide a high level of availability. Therefore, the unsuccessful map work may be rectified by reloading the replica. The unsuccessful reduce job can be retried by retrieving the data from the finished map tasks again.

\section{Programming Model for MapReduce}

MapReduce uses condensed parallel data processing approach, and runs on a cluster of computers. Its programming part comprises of two user defined functions: 1. \textit{map}, and 2. \textit{reduce}, as shown in Table~\ref{tab:mred1}. The map function is provided with a collection of key/value pairs of data as its inputs. On submission of a MapReduce job to the  system, map tasks (called mappers) are initiated on the compute nodes. Each map task applies the map function to every key/value pair \textit{(k1, v1)}. For a given input key/value combination it is possible to construct zero or more intermediate key/value pairs in the form of a list \textit{(k2, v2)}. These interim results are sorted by the keys and saved in the local file system in a distributed file system.

\begin{table}[!ht]
\centering
\caption{The MapReduce functions: \textit{Map} and \textit{Reduce}}
\label{tab:mred1}
\begin{tabular}{l|l}
\hline
Function & Arguments\\
\hline
\textit{map}   & \textit{(k1, v1)} $\to$ \textit{list(k2, v2)}\\
\hline
\textit{reduce} & \textit{(k2, list(v2))} $\to$ \textit{list(v3)}\\
\hline
\end{tabular}
\end{table}

On completion of all the map tasks, the MapReduce engine informs about the  reduce tasks to the \textit{reducer} to start processing. The reducers are also processes, and they will pickup the output files from the map tasks in parallel, and then merges these files. This will be a merge-sort as the files are already sorted. The output of map tasks is used to combine the key/value pairs into a set of new key/value pairs $(k2, list(v2))$. Here, the values with same key $k2$ are grouped into a list, and used as input to the reduce function. The reduce function makes use of a user-defined processing logic to process the data. The results, which is normally a list of values, are written back to the storage system.\medskip

\noindent
{\em Example of case study: Calculating total revenue for a website for each source IP}. As an illustrative instance, considering the database UserVisits shown as database Table~\ref{tab:uservist}, a common task involves computing the aggregate sum of revenue, referred to as \textit{revnuSum}, for each distinct \textit{source IP} address that accesses the website. The approach is as follows: add up all of the money from advertising on all of the websites that are showing this product information. The website in question has a \textit{source IP}and shows certain product information. These websites are scattered across different geographical locations. The task is performed autonomously for all websites by employing parallel and distributed method.

\begin{table}[!ht]
\footnotesize
\centering
\caption{UserVisits Table}
\label{tab:uservist}
\begin{tabular}{c|c|c|c}
\hline
ID-$>$ & sourceIP & destIP & revnuSum \\
\hline
ID Type-$>$ & char(16) & char(100)  & float\\
\hline 
\hline
ID-$>$ & userAgent & searchWord & duration \\
\hline
ID Type-$>$ & char(64) & char(32) & int \\
\hline
\end{tabular}
\end{table}

A typical MapReduce algorithms for its \textit{map} and \textit{reduce} functions are given as Algorithms~\ref{algo:mapfvst} and~\ref{algo:redfvst}, respectively.

\begin{algorithm}[!ht]
\caption{Map Function for Users Visits} 
\label{algo:mapfvst}
\begin{algorithmic}[1]
\STATE \textbf{input}: charString key, value 
\STATE charString[] array =  value.split(``$|$");
\STATE Emit\_Intermediate(array[0], ParseFloat (array [2]);    
\end{algorithmic}
\end{algorithm}

\begin{algorithm}[!ht]
\caption{Reduce Function for UserVisits} 
\label{algo:redfvst}
\begin{algorithmic}[1]
\STATE \textbf{input}: charString key, Iterator values 
\STATE float revnue\_Sum = 0;
\WHILE{values.Next()!= NULL}
\STATE $|$ revnue\_Sum += values.next();
\ENDWHILE
\STATE Emit(key, revnue\_Sum); 
\end{algorithmic}
\end{algorithm}

It is assumed that that the input data set is in text format, the tuples separated by lines, and columns are separated by the separator character ''$|$". Every mapper in the system parses the tuples that are assigned to it and produces a key/value pair in the form of (sourceIP, revnuSum). The outcomes are initially stored in the local disc of the mapper and subsequently transferred (called shuffling) to the reducers. During the reduction phase, the key/value pairs are organized into groups denoted as (sourceIP, (revnuSum1, revnuSum2, revnuSum3, ...)), where the grouping is determined by the keys, namely the sourceIP. In order to handle these pairs, each reducer summarises the revnuSum for a single sourceIP, and the resulting value (sourceIP, sum(revnuSum)) is created and returned.\hfill $\Box$\medskip

\section{Word frequency count experiment}

\begin{figure}[!ht]
\centering
\includegraphics[scale=.15]{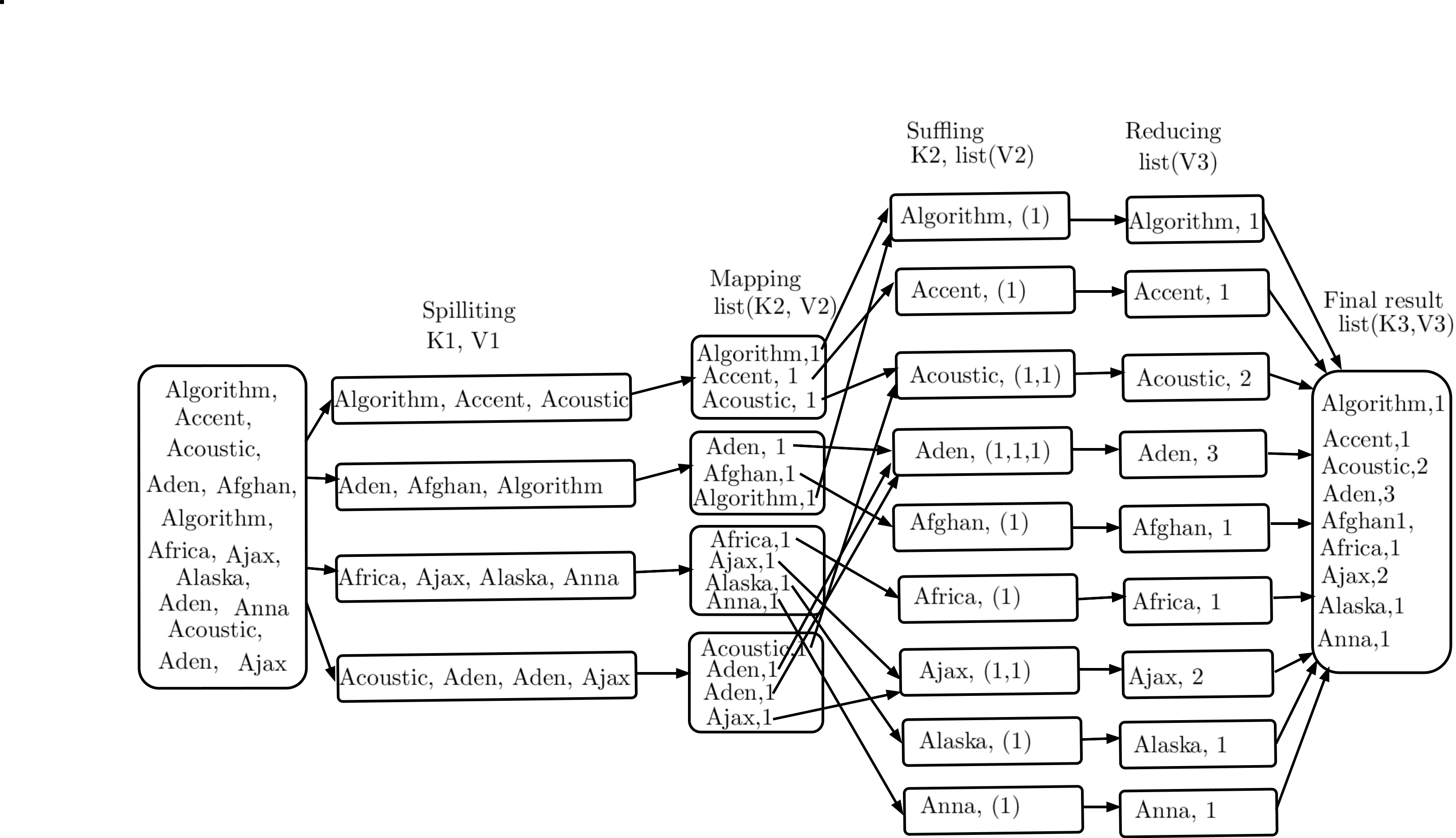}
\caption{Word frequency counting through MapReduce}
\label{fig:aggrgt}
\end{figure}

An experiment was conducted for \textit{word frequency count} for a huge database (big-data), using the MapReduce technique. The obtained results were compared with alternative methodologies. Fig.~\ref{fig:aggrgt} illustrates the detailed functioning of the system at a smaller scale. Considering that the following terms are present in the \textit{tokens.txt} file: \textit{Algorithm, Accent, ..., Ajax}. For the purpose of elucidation, we have taken a much-reduced form. The MapReduce method uses the process depicted in Fig.\ref{fig:aggrgt} to calculate the frequency of each word.

\begin{enumerate}
\item[i.] Work is first \textit{split} into four parts so that it can be done on four different nodes.

\item[ii.] Next, the \textit{mappers} tokenize their share of the job in the form of $list(K2, V2)$, such that each word is a token with count $1$.

\item[iii.] Next, \textit{sorting} and \textit{ shuffling} take place. The key $K2$ is a token, and $list(V2)$ is the combination of $V2$ from the mapper for matched tokens. The key-value pairs ($list(V3) $) are sent to the \textit{reducer,}

\item[iv.] Lastly, the key/value pairs are gathered and written to the output, which is an HDFS file. 
\end{enumerate}

\subsection{Big-data processing results}

For this experiment, three files of 350 MB, 1 GB and as well as 2 GB all included token (keyword) sequences that had no particular order; this corresponds to the first box in Fig.~\ref{fig:aggrgt}. The file \textit{tokens.txt} was utilised for each size, but with varying sizes.

For Hadoop platform, a single java program was written to run on all the nodes, in the Hadoop distributed file system (dfs). A shell script "run.sh" initially builds a Hadoop distributed file system in a directory called ``Count" to store the key-word counts, and the input database is \textit{tokens.txt.} This shell script builds the Java program, runs it on Hadoop, and then shows the working time in seconds. This information is provided in second column of table\ref{tab:wfreq2}.

\begin{table}[!ht]
\begin{center}
\caption{Word frequency counting optimum results on different platforms}
\label{tab:wfreq2} 
\begin{tabular}{|p{2cm}|p{2cm}|p{1cm}|p{1cm}|}
\hline
Platform $\rightarrow$  & Hadoop (MR) & Spark & Hive \\
File size ($\downarrow$)& (secs.) & (secs.)   &  (secs.)\\  
\hline
Small (350 MB)         &~79.355    & 11.657    & ~43.346\\
Medium (1 GB)           &216.076   & 31.112   & 125.72\\
Large (2GB)             & 416.952   & 77.065  & 218.030\\
\hline
\end{tabular} 
\end{center} 
\end{table}

\begin{figure}[!ht]
\centering
\includegraphics[scale=0.31]{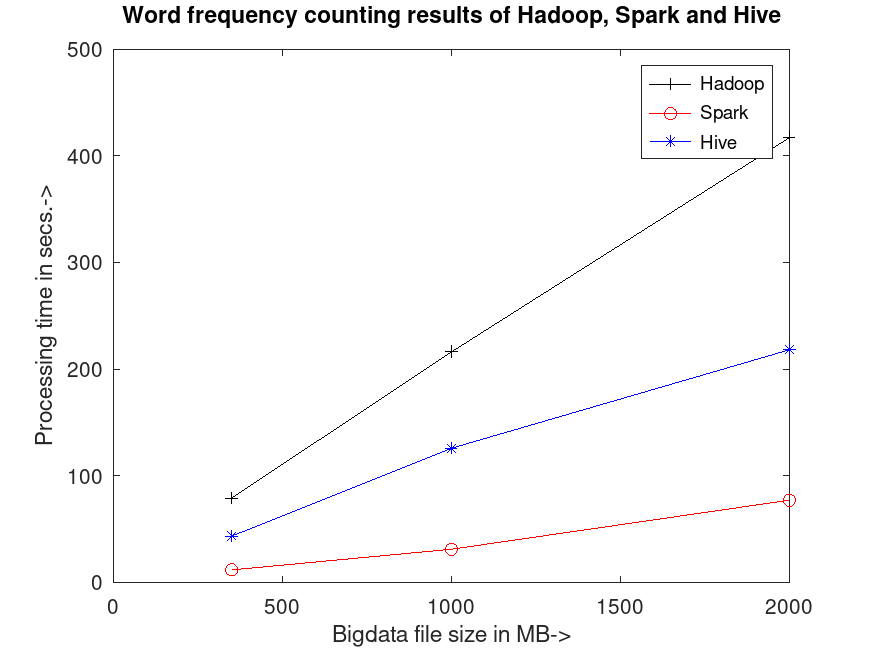}
\caption{Word frequency counting result on Hadoop, Spark, and Hive}
\label{fig:graph5}
\end{figure}

In addition, the Scala programming language was used, through its inherent mapreduce capabilities, for the purpose of calculating the frequency of words. Scala utilizes a single node as opposed to a distributed file system such as Hadoop. The Scala program that was run on Spark machine (a variant of Unix/Linux), the times for three different big-data sizes are shown in the third column of table~\ref{tab:wfreq2}, and the plots in Fig.~\ref{fig:graph5} show the relative performance of data shown in the table.

The third result was achieved by performing sql queries on Hive platform, and the result in seconds are shown in the last column of table \ref{tab:wfreq2}. 

Hadoop's use of the distributed file system (DFS) and ability to provide distributed processing, as opposed to the other two's usage of single-node programming, makes it significantly different from each of them. The increased duration of time required for executing Depth-First Search (DFS) might be caused due to delays in communications between Hadoop nodes.

\subsection{OpenMP Results}

\begin{figure}[!ht]
\centering
\includegraphics[scale=0.31]{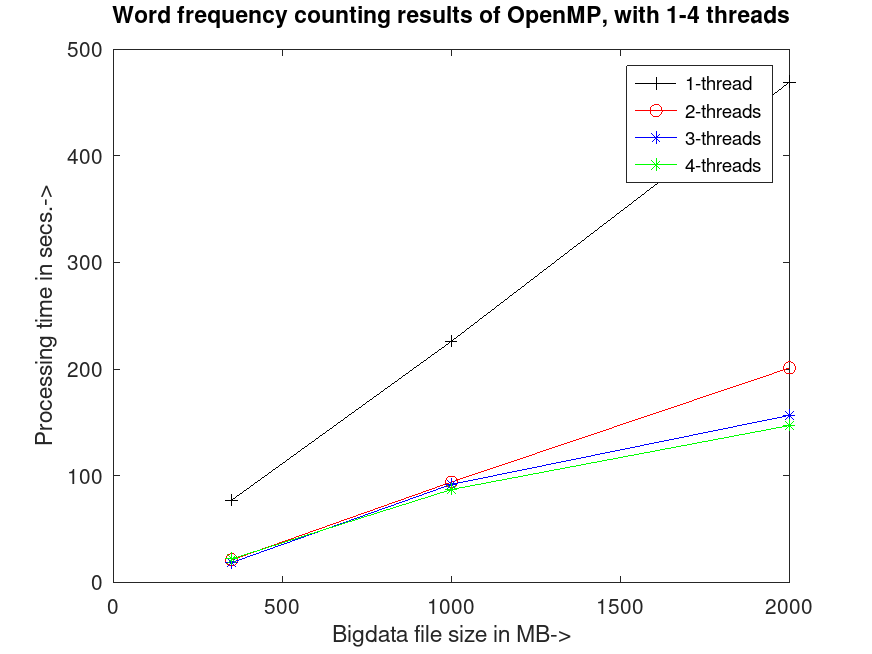}
\caption{Word frequency counting on OpenMP 4-Core Machine}
\label{fig:graph6}
\end{figure}

The word-frequency count experiment was conducted out on the parallel processing platform of OpenMP for processor threads 1, 2, 3, and 4. Database file sizes of 350 MB, 1 GB, and 2 GB were tracked for each thread value. The computation time for word count for each pair of ``thread-value $\times$ file-size" was determined. The time taken, for different database size as well as for different thread count, in seconds, for each category are presented in Table ~\ref{tab:wfreq1} and  in (Fig.~\ref{fig:graph6}). The threads are effectively the processing elements, it shows an almost linear increase in performance as the number of threads increase, however, it slightly deviates from exact linear because the processor has devote itself to do other housekeeping jobs like memory management and processing switching.

\begin{table}[!ht]
\begin{center}
\caption{Word frequency counting on OpenMP 4-Core Machine}
\label{tab:wfreq1} 
\begin{tabular}{|p{2cm}|p{1cm}|p{1cm}|p{1cm}|p{1cm}|}
\hline
Threads $\rightarrow$  & 1-thread & 2-threads & 3-threads & 4-threads\\
File size ($\downarrow$)& (secs.) & (secs.)   &  (secs.)  & (secs.)\\  
\hline
Small (350 MB)         &~77.065    & ~20.957    & ~18.485    & ~21.958\\
Medium (1 GB)           &225.945   & ~94.425    & ~91.792    & ~87.369\\
Large (2 GB)             & 469.34   & 201.428   & 156.469   & 147.031\\
\hline
\end{tabular} 
\end{center} 
\end{table}

\section{Conclusion}

We have presented the application of distributed processing using MapReduce algorithm, highlighting its simple structure, and have demonstrated its use for handling large (big) data. Its distributed working has been explained through a web application, and corresponding algorithms, where an advertisement agency bills from a product making company based how many times there has been mouse clicks on  all the advertisements running in geographically distributed manner on in distributed physical locations.

Apart from presenting its working in simple language, while comparing it with the traditional distributed processing and lattersits drawbacks, it has shown that this algorithm is scalable by increasing the size of big data from 350 MB to 2000 MB, with no degradation in performance. The truly distributed processing system Hadoop where MapReduce is running, shows a almost 100 percent linear response as a function of size of big data.

\onecolumn 
 
\end{document}